\documentclass[a4paper,11pt]{article}
\pdfoutput=1 

\usepackage{jcappub} 

\usepackage{multicol}
\usepackage{multirow}
\usepackage[T1]{fontenc} 
\usepackage{graphicx}
\usepackage[export]{adjustbox}
\usepackage{amsmath}
\usepackage{array}
\usepackage{subcaption}
\usepackage{amsfonts}
\usepackage{amssymb}
\usepackage[left=2cm,right=2cm,top=2cm,bottom=2cm]{geometry}
\usepackage{array}
\usepackage{multirow}
\usepackage{multicol}
\usepackage{tablefootnote}
\usepackage{hyperref}
\usepackage[normalem]{ulem} 
\usepackage{chngcntr}
\usepackage{scalerel}
\usepackage{xcolor} 
\definecolor{DeepGreen}{RGB}{0,150,150} 

\title{\boldmath An information-theoretic framework for quantifying coherence in the cosmic density field}

\author[a]{Biswajit Pandey}

\author[a]{and Anindita Nandi}

\affiliation[a]{Department of Physics, Visva-Bharati University, Santiniketan, 731235, India}

\emailAdd{biswap@visva-bharati.ac.in} \emailAdd{anindita.nandi96@gmail.com}

\abstract{We introduce an information-theoretic framework for
  quantifying the coherence retained in the cosmic density field
  across cosmic time. Using mutual information between density fields
  at different redshifts in cosmological $N$-body simulations, and
  subtracting an appropriate reference baseline, we isolate the excess
  statistical dependence generated by gravitational evolution. Unlike
  conventional clustering statistics, this approach is sensitive to
  nonlinear and higher-order correlations and directly measures the
  shared information between different cosmic epochs. The shared
  mutual information exhibits a strong dependence on spatial scale,
  increasing as the density field is progressively coarse-grained,
  reaching a maximum, and declining at larger scales. Through a series
  of tests involving different simulation volumes, discretization
  schemes, alternative baseline constructions, and subvolume analyses,
  we show that the location of this maximum is governed primarily by
  finite-sampling effects in the mutual information estimator rather
  than by a fundamental physical coherence scale. The observed peak
  therefore provides an empirical indication of the range over which
  the present estimator yields robust measurements. We further
  demonstrate that the mutual information is sensitive to cosmic
  evolution and cosmological parameters. Applying the framework to the
  cosmic web, we find that filamentary environments retain more
  statistically accessible shared information between cosmic epochs
  than sheets and clusters, with this trend remaining robust after
  controlling for sample size. We also note that the shared
  information between the initial and present-day density fields is
  found to be positive but extremely small, reflecting the progressive
  redistribution of information during nonlinear structure
  formation. Our results establish mutual information as a powerful
  new tool for quantifying coherence, memory, and information
  retention in the large-scale structure of the Universe.}

\begin{document}
\maketitle
\flushbottom

\section{Introduction}

The large-scale structure of the Universe is the cumulative outcome of
gravitational instability acting on primordial density fluctuations
over cosmic time. Galaxies, groups, clusters, filaments, and walls
form a complex cosmic web whose morphology encodes information about
both the initial conditions of the Universe and the subsequent history
of gravitational evolution. A central challenge in modern cosmology is
to quantify how much of this information is retained in the evolved
density field and on what spatial scales the matter distribution
remains coherent across cosmic time. This question underlies a broad
range of problems, including the emergence of large-scale structure,
the transition to statistical homogeneity, environmental influences on
galaxy formation, and the interpretation of cosmological simulations
and surveys.

Observational studies of galaxy clustering have revealed coherent
structures extending over tens of megaparsecs while also indicating a
gradual approach to homogeneity on larger scales
\citep{bharadwaj04, yadav05, hogg05, sarkar09, pandey11,
scrimgeour12, nadathur13, pandey15, sarkar16, sarkar19,
sarkar23}. Numerical simulations have played a crucial role in
interpreting these observations and in exploring the growth of cosmic
structure across a wide range of scales
\citep{springel05, kim11, schaye15, nelson19}. Despite remarkable
progress, most analyses remain rooted in correlation functions, power
spectra, and related clustering statistics, which quantify the
strength of fluctuations but do not directly measure how much
information about one epoch is retained in another.

Gravitational evolution is intrinsically nonlinear and nonlocal.
Processes such as mode coupling, bulk flows, mergers, and tidal
interactions redistribute information across scales and spatial
locations, making the concept of coherence fundamentally an
information-theoretic one. A natural question therefore arises: how
much information about the cosmic density field at one epoch remains
encoded in the density field at a later epoch, and how can this be
quantified in a statistically rigorous manner?

Mutual information (MI) provides a natural framework for addressing
this question. Unlike conventional two-point statistics, MI captures
all statistical dependencies between two fields, including nonlinear
and higher-order correlations, without assuming Gaussianity or linear
evolution. In the context of large-scale structure, it measures how
much knowledge of the density field at one epoch reduces uncertainty
about the density field at another. Information-theoretic approaches
have been applied to a variety of cosmological problems
\citep{hosoya04, pandey13, hall13, pandey16a, pandey16b, vazza17,
pandey17a, pandey17b, pandey21, pandey23, moon24, nandi24}, but
their potential as a quantitative probe of coherence in the evolving
cosmic density field remains largely unexplored.

\begin{figure*}
\centering
\includegraphics[width=\columnwidth]{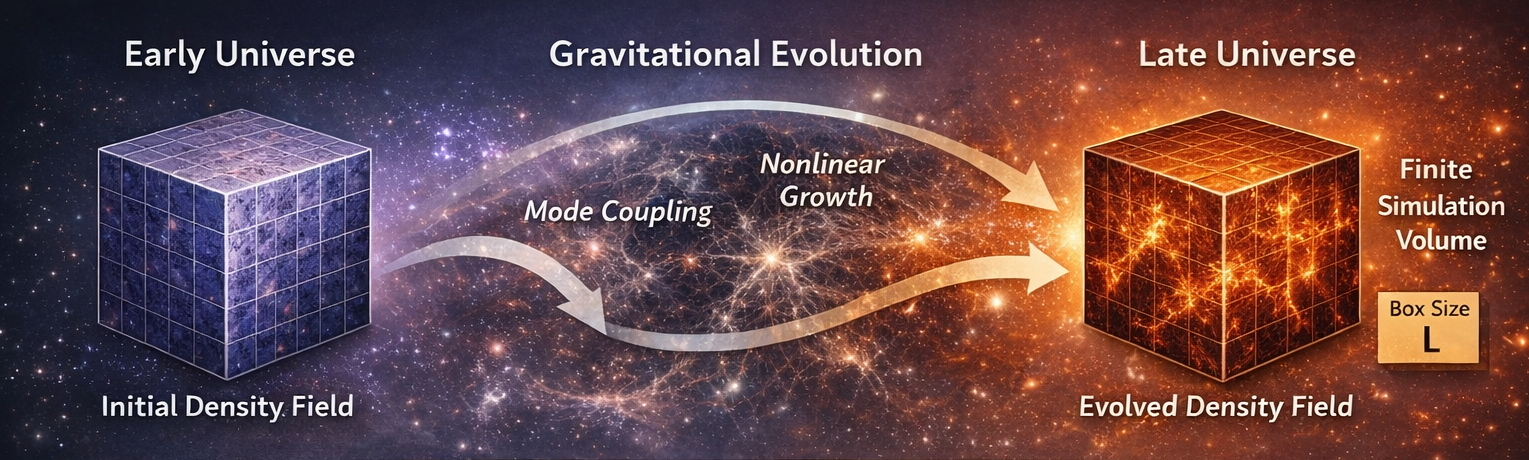}
\caption{The same comoving region of the Universe is evolved from an
  early epoch (left) to a later epoch (right) through nonlinear
  gravitational dynamics. While gravity redistributes information via
  mode coupling, bulk flows, and nonlinear growth, the finite
  simulation volume of side length $L$ limits the longest wavelengths
  that can be retained. Our analysis quantifies how much information
  about the early-time density field remains encoded in the evolved
  field by treating gravitational evolution as an information channel
  and measuring the shared mutual information between snapshots.}
\label{fig:zigu0}
\end{figure*}

In this work, we develop an information-theoretic framework for
measuring coherence in the cosmic density field using mutual
information. The conceptual framework is illustrated in
\autoref{fig:zigu0}, where gravitational evolution connects the
density fields of the same comoving region at two epochs, and mutual
information quantifies their shared statistical dependence. We
quantify the information shared between density fields at different
cosmic epochs and introduce a physically motivated baseline
subtraction that isolates the component associated with gravitational
evolution. This construction enables us to study how shared
information varies with spatial scale, redshift separation, and
cosmological parameters, thereby providing a new perspective on the
retention and redistribution of information during structure
formation.

An equally important aspect of this study is the assessment of the
estimator itself. Mutual information estimates depend on finite
sampling, voxelization, and probability estimation, all of which can
introduce scale-dependent biases. By performing a series of robustness
tests involving different simulation volumes, subvolumes,
discretization schemes, baseline constructions, and redshift
separations, we investigate the reliability and limitations of the
proposed framework. We find that while the method successfully traces
the evolution of shared information in the density field, an apparent
characteristic scale emerges that is primarily driven by estimator and
sampling effects rather than by a fundamental physical coherence
scale. This behaviour provides a useful empirical guide to the range
over which the present estimator can be reliably interpreted.

The framework developed here is intended not as a replacement for
traditional clustering statistics, but as a complementary tool for
quantifying coherence, memory, and information retention in nonlinear
gravitational systems. By recasting these questions in
information-theoretic terms, our approach opens a new avenue for
studying the evolution of cosmic structure and for assessing the
strengths and limitations of information-based analyses in
cosmological datasets.

The paper is organized as follows. Section~\ref{sec:data_method}
describes the simulations and the information-theoretic methodology.
Section~\ref{sec:results} presents the measurements of shared mutual
information and examines their robustness to various analysis choices.
Finally, Section~\ref{sec:conclusion} summarizes our findings and
discusses their implications for future applications of
information-theoretic methods in cosmology.

\section{Data and Method of Analysis}
\label{sec:data_method}

In this section, we describe the numerical data used in this work and
outline the information-theoretic framework employed to quantify the
retention of gravitational information across cosmic time. Our aim is
to develop a method that captures the full statistical dependence
between density fields at different epochs while cleanly separating
physical correlations generated by gravitational evolution from
spurious correlations arising due to finite sampling, discretization,
and finite simulation volume.

\subsection{Simulation Data}
\label{subsec:data}
We simulate the evolution of the cosmic density field at redshifts
$z=2$, $z=1$ and $z=0$ in a $\Lambda$CDM cosmology using parameters
consistent with the \emph{Planck} measurements \cite{aghanim20} in a
standard $\Lambda$CDM cosmology. We consider simulation volumes with
side lengths $L=512~\mathrm{Mpc}$, $256~\mathrm{Mpc}$, and
$128~\mathrm{Mpc}$ to systematically assess the impact of finite
volume on the retained gravitational information. For each box size
$L$, we employ a particle-mesh setup with $N=(L/2)^3$ particles
evolved on a $(L)^3$ mesh (with lengths in Mpc), ensuring that both
mass and force resolution scale consistently across all simulations.
The simulations are performed using an open-source $N$-body code based
on the particle-mesh (PM) technique \cite{bharad04a, mondal15}, which
efficiently captures the large-scale gravitational dynamics relevant
for the present analysis.  The code is publicly
available \footnote{\url{https://github.com/rajeshmondal18/N-body}}.
For each simulation box size considered in this work, we generate 10
independent realizations of the initial density field. These
realizations are used to estimate the $1\sigma$ uncertainties in our
measurements. We analyze the snapshots from the $N$-body
simulations. Each simulation occupies a cubic comoving volume of side
length $L$ with periodic boundary conditions and contains a fixed
number of collisionless dark-matter particles. Snapshots at different
redshifts correspond to the same comoving volume, enabling a direct,
voxel by voxel comparison of the evolving cosmic density field.

\subsection{Method of Analysis}
\label{subsec:method}

A fundamental challenge in studies of cosmic structure formation is to
quantify how much information about the density field at one epoch is
retained at another. Conventional statistical measures, such as the
two-point correlation function and power spectrum, characterize the
amplitude of density fluctuations but do not directly quantify the
amount of information shared between density fields at different
times. Moreover, nonlinear gravitational evolution couples modes
across scales and generates higher-order correlations that are not
fully captured by second-order statistics alone.

To address this problem, we adopt an information-theoretic framework
based on mutual information (MI), which provides a direct and
model-independent measure of the information shared between two random
fields. Unlike correlation-based statistics, MI is sensitive to all
statistical dependencies, including nonlinear and higher-order
relationships. It therefore offers a natural framework for quantifying
the degree of coherence retained by the cosmic density field across
cosmic time.

We consider two snapshots of an $N$-body simulation at redshifts
$z_1$ and $z_2$, occupying the same comoving cubic volume of side
length $L$. The simulation volume is partitioned into cubic voxels of
side length $d$, yielding

\begin{equation}
N_{\rm vox}=\left(\frac{L}{d}\right)^3
\end{equation}

voxels. For each voxel $i$, we define the discrete random variables

\begin{equation}
X_i \equiv n_i(z_1), \qquad
Y_i \equiv n_i(z_2),
\end{equation}

where $n_i(z)$ denotes the number of simulation particles contained
within voxel $i$ at redshift $z$. The sets $\{X_i\}$ and $\{Y_i\}$
therefore represent coarse-grained realizations of the matter density
field at the two epochs, viewed at a spatial resolution $d$.

To estimate the underlying probability distributions, the voxel counts
are grouped into $n_{\rm bin}$ density bins. The marginal probability
that a voxel belongs to density bin $a$ at redshift $z_1$ is

\begin{equation}
p_{\scaleto{X}{3.5pt}}(a)=\frac{N_a}{N_{\rm vox}},
\end{equation}

where $N_a$ is the number of voxels assigned to bin $a$. An analogous
construction yields the marginal distribution
$p_{\scaleto{Y}{3.5pt}}(b)$ for the snapshot at redshift $z_2$.

The joint probability distribution
$p_{\scaleto{XY}{3.5pt}}(a,b)$ is obtained by pairing voxels at the
same spatial locations in the two snapshots and counting their joint
occurrences in density bins $(a,b)$. While the marginal distributions
describe the density statistics at individual epochs, the joint
distribution encodes the statistical relationship between the two
density fields and therefore contains information about the coherence
retained during gravitational evolution.

The mutual information between the two density fields is then given by
\citep{shannon48}

\begin{equation}
I_{\rm data}(X;Y)=
\sum_{a=1}^{n_{\rm bin}}
\sum_{b=1}^{n_{\rm bin}}
p_{\scaleto{XY}{3.5pt}}(a,b)
\log
\left[
\frac{p_{\scaleto{XY}{3.5pt}}(a,b)}
     {p_{\scaleto{X}{3.5pt}}(a)\,
      p_{\scaleto{Y}{3.5pt}}(b)}
\right].
\label{eq:mi}
\end{equation}

Mutual information measures the reduction in uncertainty about one
field obtained from knowledge of the other. It vanishes only when the
two fields are statistically independent and increases as their shared
information grows. In the present context, $I_{\rm data}(X;Y)$
quantifies the degree to which the density field at one epoch remains
predictive of the density field at another epoch.

An important advantage of MI is that it does not assume Gaussianity,
linearity, or any specific physical model. It is sensitive to the full
probability distribution and therefore captures information encoded in
higher-order correlations that are inaccessible to the power spectrum
alone. Unlike the power spectrum, which characterizes scale-dependent
spatial clustering within a single density field, the mutual
information considered here quantifies the statistical dependence
between density values at matched spatial locations at different
epochs. Because mutual information is sensitive to the full joint
probability distribution, it can capture nonlinear statistical
dependence between the two fields, providing information complementary
to that contained in two-point clustering statistics.  The measure
proposed in this work provides a direct probe of how gravitational
evolution preserves, redistributes, or erases information in the
cosmic density field.

Because mutual information is symmetric, $I_{\rm data}(X;Y)=I_{\rm
  data}(Y;X)$, the present analysis measures shared coherence between
epochs rather than a directional flow of information. By evaluating
$I_{\rm data}(X;Y)$ over a range of voxel sizes $d$, we investigate
how the information shared between density fields depends on spatial
scale and how reliably such coherence can be inferred from
finite-volume simulations.

The present analysis uses particle counts within cubic voxels as the
coarse-grained density variables. The information-theoretic framework
itself is not restricted to this particular density-assignment scheme
and can, in principle, be formulated using standard mass-assignment
methods such as cloud-in-cell (CIC). In such a formulation, the CIC
density estimates at the regular grid locations could be used directly
as the variables entering the mutual information estimator. A precise
comparison with the present voxel-count formulation would, however,
require a consistent prescription for the associated spatial sampling
and smoothing, since averaging CIC values to obtain voxel densities
would introduce an additional coarse-graining step and thereby define
a different estimator. A systematic comparison of particle-count and
CIC-based mutual information measurements is therefore left for future
work.

\subsection{Information-theoretic estimator}

Mutual information provides a natural measure of the statistical
dependence between the density fields at two cosmic epochs. In
addition to the probability-based expression in
\autoref{eq:mi}, it can be written in the equivalent entropy form
\begin{equation}
I(X;Y)=H(X)+H(Y)-H(X,Y),
\end{equation}
where $H(X)$ and $H(Y)$ are the Shannon entropies of the marginal
density distributions at redshifts $z_1$ and $z_2$, respectively, and
$H(X,Y)$ is their joint entropy. This representation makes the
information-theoretic interpretation particularly transparent:
mutual information measures how much the uncertainty in one density
field is reduced by knowledge of the other. In the context of cosmic
structure formation, it quantifies the amount of information that
remains shared between two stages of gravitational evolution.

Estimating information-theoretic quantities from finite data is,
however, inherently challenging. Entropy and mutual information
estimators are known to suffer from finite-sample biases arising from
the incomplete sampling of the underlying probability distributions
\citep{miller55, paninski03}. In our analysis, this issue becomes
increasingly important at large voxel sizes, where the number of
available voxels decreases rapidly and the probability distributions
are inferred from a progressively smaller number of independent
samples. Accounting for such effects is therefore essential for a
reliable interpretation of the measured scale dependence.

To reduce the leading-order finite-sample bias, we employ the
Miller--Madow correction \citep{miller55, paninski03}. For a discrete
distribution with $K$ occupied bins estimated from $N$ samples, the
corrected entropy is
\begin{equation}
H_{\rm MM}
=
H_{\rm naive}
+
\frac{K-1}{2N\ln 2},
\end{equation}
where $H_{\rm naive}$ is the standard plug-in entropy estimator. All
mutual information measurements presented in this work are computed
using these corrected entropy estimates. This procedure substantially
reduces artificial contributions arising from finite sampling and
provides a more robust characterization of the shared information
between density fields.

Even after applying entropy-bias corrections, finite sampling and
discretization can generate a non-zero mutual information signal in
the absence of any physical correlation. To isolate the contribution
associated with gravitational evolution, we introduce the shared
physical mutual information
\begin{equation}
I_{\rm phys}(X;Y)
=
I_{\rm data}(X;Y)-I_{\rm null}(d),
\label{eq:smi}
\end{equation}
where $I_{\rm data}(X;Y)$ is the mutual information measured from the
simulation snapshots and $I_{\rm null}(d)$ is defined as the mutual
information computed using the same voxel size $d$ between two
independent, structureless random point distributions with identical
particle number and box size. The null term quantifies the residual
information expected from finite sampling and discretization alone and
therefore serves as a baseline against which physically meaningful
correlations can be assessed.

The quantity $I_{\rm phys}(X;Y)$ thus provides a measure of the
information retained by the density field across cosmic time beyond
that expected from statistical fluctuations. In other words, it
quantifies the degree to which knowledge of the density field at one
epoch improves our ability to predict the density field at another
epoch. Throughout this work, we use the scale dependence of
$I_{\rm phys}(X;Y)$ as a diagnostic of how gravitational evolution
preserves, redistributes, and erases information across different
spatial scales.

A key objective of our analysis is not only to measure the shared
information itself, but also to assess the robustness of the estimator
used to infer it. We therefore systematically examine the dependence
of the results on the number of density bins, simulation volume,
redshift separation, and the choice of null baseline. These tests
allow us to distinguish genuine physical trends from estimator-driven
effects and to identify the range of scales over which the measured
mutual information can be interpreted reliably.

\begin{figure*}[t]
    \centering
    \includegraphics[width=\columnwidth]{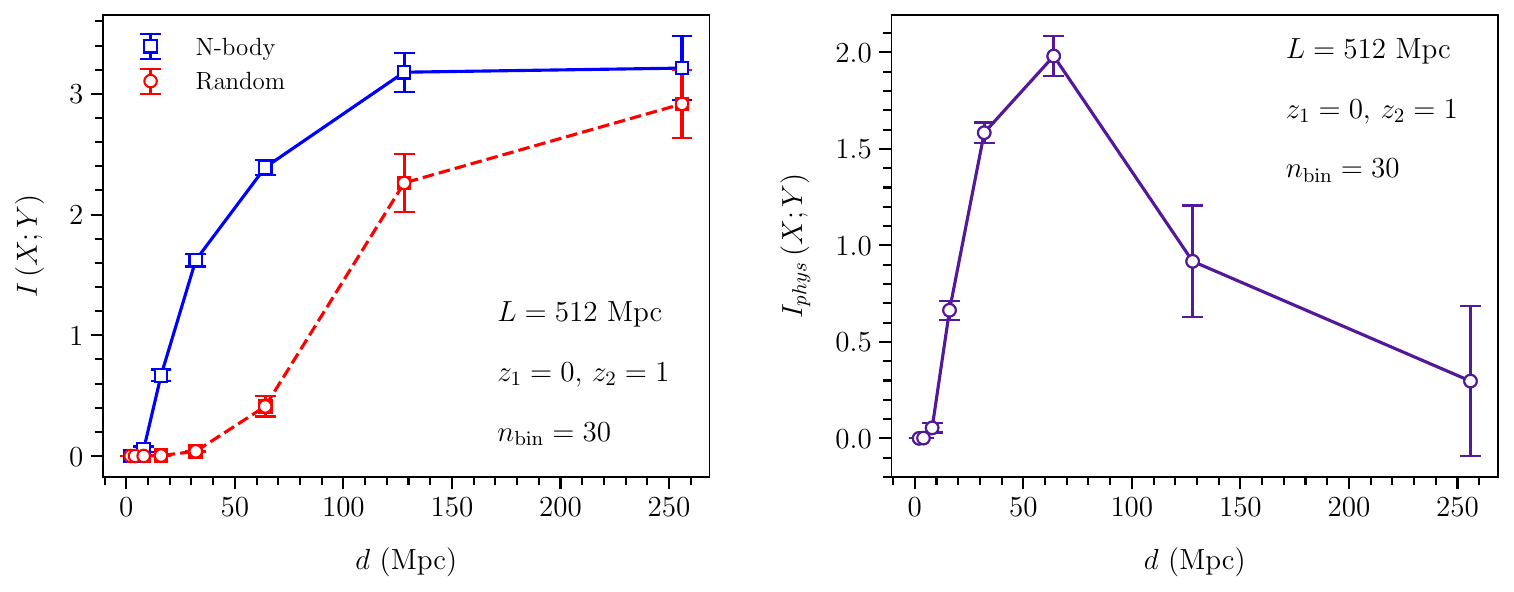}
    \caption{ \textbf{Scale dependence of the shared mutual
        information between cosmic density fields at $z_1=0$ and
        $z_2=1$.}  \emph{Left:} Mutual information measured from the
      $N$-body density field, $I_{\rm data}(X;Y)$, together with the
      corresponding estimate from structureless random catalogs,
      $I_{\rm null}(d)$, which quantifies contributions arising from
      finite sampling and discretization in the absence of
      gravitationally generated structure.  \emph{Right:} Shared
      physical mutual information, $I_{\rm phys}(X;Y)=I_{\rm
        data}(X;Y)-I_{\rm null}(d)$, isolating the information
      associated with gravitational evolution.  The shared information
      exhibits a strong dependence on the voxel size $d$: it is weak
      on small scales, increases with coarse-graining, and reaches a
      characteristic maximum before decreasing at larger scales.  This
      behaviour demonstrates that the amount of information shared
      between density fields at different cosmic epochs depends
      sensitively on the spatial scale at which the field is
      sampled. Subsequent analyses presented in this work show that
      the location of the maximum is influenced primarily by
      sample-size and estimator effects, motivating a careful
      assessment of the reliability range of mutual information-based
      coherence measurements.  Error bars indicate $1\sigma$
      uncertainties estimated from 10 independent realizations of the
      $N$-body simulations and random catalogs.  }
    \label{fig:zigu1}
\end{figure*}

\section{Results}
\label{sec:results}
In this section, we explore how mutual information can be used to
quantify coherence in the cosmic density field across cosmic time. We
begin by examining the scale dependence of the shared information
between density fields at different epochs and then assess the impact
of voxelization, sample size, simulation volume, and baseline
construction on the inferred signal. These tests provide a systematic
characterization of the estimator and its range of reliability.
Finally, we investigate the dependence of the measured coherence on
cosmic evolution and cosmological parameters, highlighting the broader
potential of this information-theoretic framework for studies of
large-scale structure.

\subsection{Measuring coherence across cosmic time}

We begin by applying our information-theoretic framework to quantify
the amount of information shared between the cosmic density field at
redshifts $z_1=0$ and $z_2=1$. \autoref{fig:zigu1} shows the mutual
information as a function of the voxel size $d$ for a simulation box
of side length $L=512~{\rm Mpc}$. The left panel presents the raw
mutual information measured from the simulation snapshots,
$I_{\rm data}(X;Y)$, together with the corresponding null estimate,
$I_{\rm null}(d)$, obtained from random realizations. The right panel
shows the shared physical mutual information,
$I_{\rm phys}(X;Y)=I_{\rm data}(X;Y)-I_{\rm null}(d)$, which isolates
the information associated with gravitationally evolved structure.

Several noteworthy features emerge. On small scales
($d \lesssim 8~{\rm Mpc}$), the shared physical mutual information is
consistent with zero within the uncertainties. At these scales,
nonlinear evolution redistributes matter efficiently through mode
coupling, mergers, and bulk flows, reducing the similarity between the
density field at different epochs when viewed at fixed Eulerian
locations. As the voxel size increases, $I_{\rm phys}(X;Y)$ rises
rapidly, indicating that progressively larger-scale features retain a
greater degree of shared information across cosmic time.

The measured signal exhibits a pronounced maximum near
$d \simeq 64~{\rm Mpc}$, followed by a gradual decline at larger
scales. At face value, this behaviour suggests that the amount of
shared information inferred from the density field depends strongly on
the scale at which the field is coarse-grained. However, the physical
interpretation of the observed maximum is not immediately obvious.
Because the number of available voxels decreases rapidly with
increasing $d$, finite-sample effects can influence the performance of
the mutual information estimator. Consequently, the peak cannot by
itself be interpreted as evidence for a preferred physical coherence
scale. Instead, it serves as a useful empirical feature whose origin
must be carefully investigated through a series of robustness tests.

The remainder of this section is devoted to this assessment. By
examining the dependence of the measured signal on simulation volume,
redshift separation, discretization choices, and alternative null
baselines, we determine which aspects of the observed scale dependence
reflect genuine properties of the evolving density field and which
arise from limitations of the estimator itself.

\begin{figure}[t]
    \centering
    \includegraphics[width=\columnwidth/2]{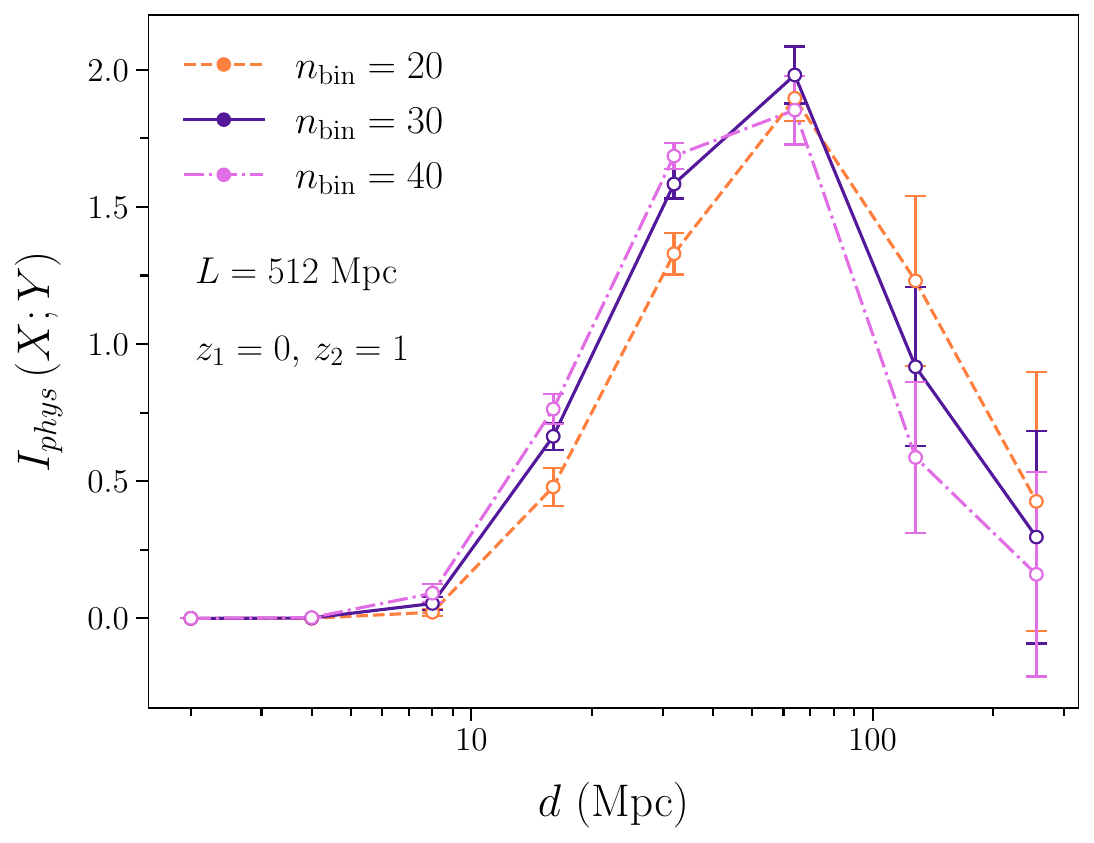}
    \caption{\textbf{Sensitivity of the shared physical mutual
        information to density binning.} The shared physical mutual
      information, $I_{\rm phys}(X;Y)$, is shown for a fixed
      simulation volume using different numbers of density bins in the
      probability estimation.  The overall scale dependence remains
      remarkably stable across all binning choices: the shared
      information increases from small voxel sizes, reaches a maximum
      at intermediate scales, and declines at larger voxel sizes. The
      persistence of this behaviour demonstrates that the measured
      trends are not driven by the specific discretization adopted for
      the density field. In particular, the location of the maximum is
      largely insensitive to the number of density bins, indicating
      that it arises primarily from the interplay between voxelization
      and the finite sample size available to the mutual information
      estimator, rather than from the details of histogram
      construction.}
    \label{fig:zigu3}
\end{figure}

\begin{figure}[t]
\centering
\includegraphics[width=\columnwidth/2]{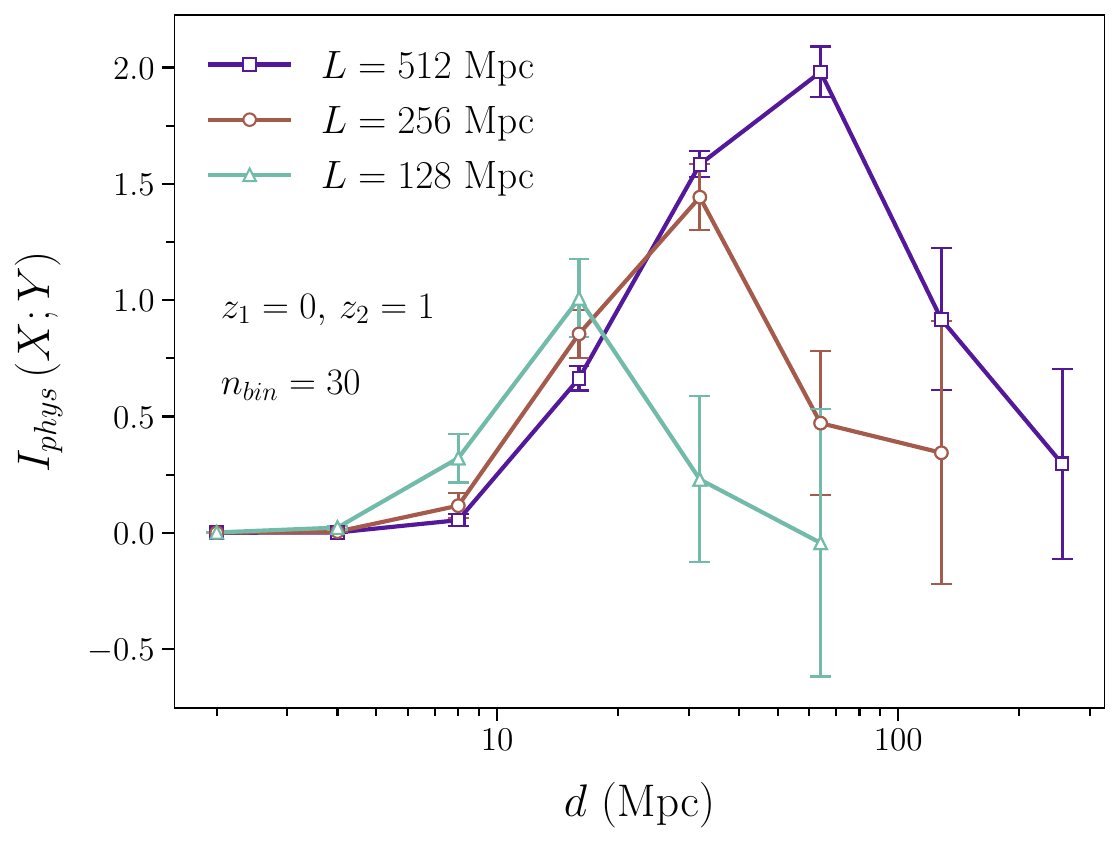}
\caption{ \textbf{Dependence of the measured shared physical mutual
    information on simulation volume.}  The shared physical mutual
  information, $I_{\rm phys}(X;Y)$, is shown as a function of voxel
  size $d$ for simulations with box sizes $L=128$, $256$, and
  $512~{\rm Mpc}$.  In all cases, the signal rises from small scales,
  reaches a pronounced maximum, and subsequently declines at larger
  scales.  The location of the maximum shifts systematically with the
  box size, occurring near $d \simeq L/8$ for each simulation volume.
  Although this behaviour initially suggests a connection with
  finite-volume effects, the dependence on $L$ alone does not
  establish a physical origin for the peak because the number of
  voxels available to the mutual information estimator also changes
  with box size.  The origin of this scaling is examined further using
  subvolume tests in \autoref{fig:new1}.}
\label{fig:zigu2}
\end{figure}

\begin{figure}[t]
    \centering
    \includegraphics[width=\columnwidth/2]{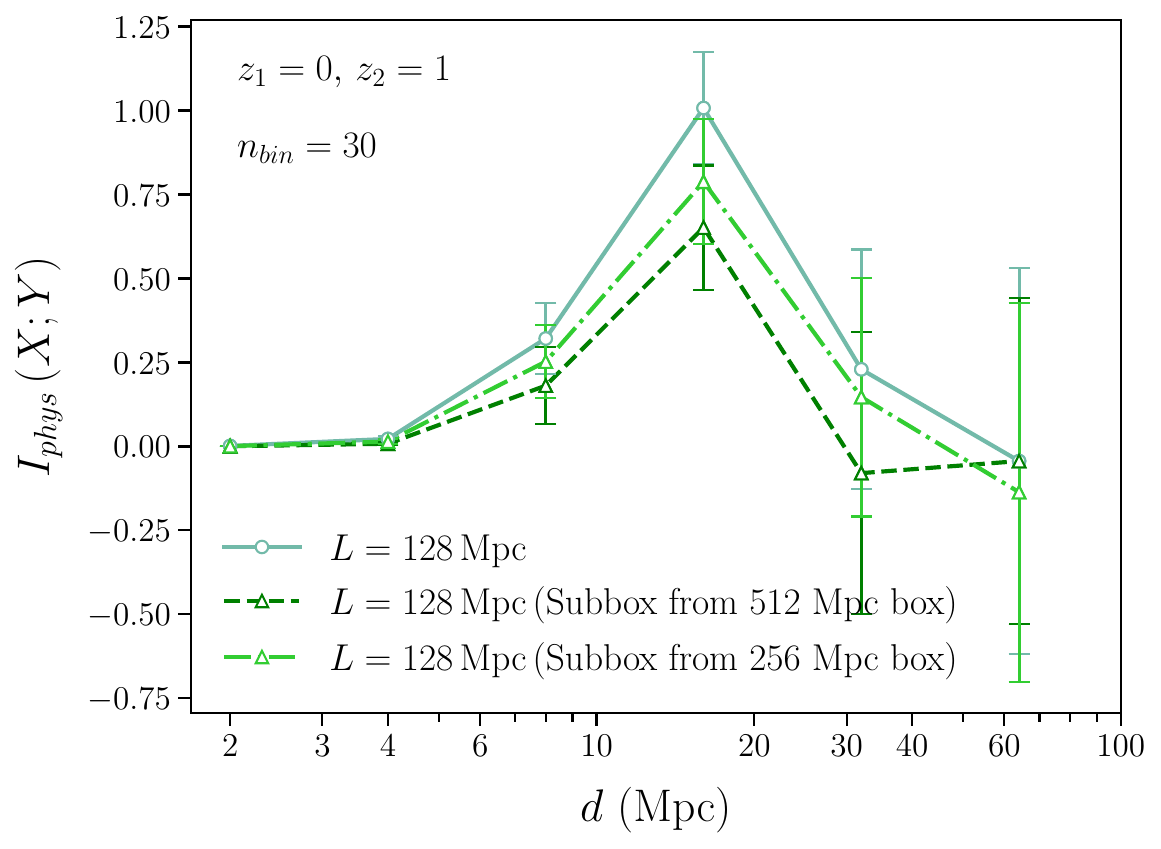}
    \caption{\textbf{Subvolume test of the mutual information
        estimator.}  Shared physical mutual information, $I_{\rm
        phys}(X;Y)$, measured for $128~{\rm Mpc}$ regions in three
      cases: an independent $128~{\rm Mpc}$ simulation, a $128~{\rm
        Mpc}$ subvolume extracted from a $256~{\rm Mpc}$ simulation,
      and a $128~{\rm Mpc}$ subvolume extracted from a $512~{\rm Mpc}$
      simulation. All three datasets therefore have the same analysed
      volume and the same number of voxels for a given voxel size $d$,
      while the latter two retain fluctuations inherited from larger
      parent volumes. The measured mutual information curves show
      broadly similar scale dependence and comparable locations of the
      characteristic maximum, with no systematic enhancement
      associated with the size of the parent simulation. This
      indicates that the behaviour near the maximum is governed
      predominantly by the effective sample size and finite-sampling
      properties of the estimator rather than by the long-wavelength
      modes present in the parent volume.}
    \label{fig:new1}
\end{figure}

\subsection{Assessing the robustness of the mutual information estimator}

A central question raised by the measurements presented in the previous
subsection is whether the observed scale dependence of the shared
physical mutual information reflects a genuine property of the evolving
density field or arises from limitations of the estimator itself. Since
the mutual information is inferred from discretized probability
distributions constructed using a finite number of voxels, both the
voxelization scheme and the available sample size can influence the
measured signal. We therefore perform a series of robustness tests to
quantify the sensitivity of our results to these effects.

We first examine the dependence on the number of density bins used to
estimate the marginal and joint probability distributions.
\autoref{fig:zigu3} shows the shared physical mutual information
measured using different binning schemes. The overall behaviour remains
remarkably stable: the signal rises from small scales, reaches a
pronounced maximum at intermediate scales, and subsequently declines at
larger scales. Although the absolute amplitude exhibits minor
variations, the qualitative form of the curve is unchanged. This
demonstrates that the measured scale dependence is not driven by the
particular choice of histogram discretization and is therefore robust
against reasonable changes in the probability-estimation procedure.

We next investigate the dependence on simulation volume.
\autoref{fig:zigu2} compares measurements obtained from simulations
with box sizes $L=128$, $256$, and $512~{\rm Mpc}$. In each case the
shared physical mutual information exhibits a similar non-monotonic
behaviour, but the location of the maximum shifts systematically with
the box size. The peak occurs near $d\simeq16~{\rm Mpc}$ for
$L=128~{\rm Mpc}$, $d\simeq32~{\rm Mpc}$ for $L=256~{\rm Mpc}$, and
$d\simeq64~{\rm Mpc}$ for $L=512~{\rm Mpc}$. At first sight, this
trend could be interpreted as evidence that the characteristic scale
is linked to the finite simulation volume through the availability of
long-wavelength modes. However, the box-size dependence alone is not
sufficient to establish such an interpretation because changing $L$
also changes the number of voxels available to the estimator at a
given coarse-graining scale.

To determine whether the behaviour near the maximum could instead
reflect genuine physical differences caused by the missing
long-wavelength modes of smaller simulation volumes, we performed a
more stringent controlled test. We extracted $128~{\rm Mpc}$ subboxes
from both the $256~{\rm Mpc}$ and $512~{\rm Mpc}$ simulations and
compared them with an independent $128~{\rm Mpc}$ simulation, applying
identical voxelization and mutual information estimation to all three
datasets. This comparison is shown in \autoref{fig:new1}. Because all
three analysed volumes have the same effective size and therefore the
same number of voxels at a given $d$, the test isolates whether
information inherited from the larger parent volumes produces a
measurable enhancement of the mutual information signal.

The three measurements exhibit broadly similar scale dependence, with
their maxima occurring near $d\simeq16\,{\rm Mpc}$, corresponding to
approximately one-eighth of the effective box size. However, the
$128\,{\rm Mpc}$ subvolumes extracted from the $256$ and $512\,{\rm
  Mpc}$ simulations do not exhibit an enhanced mutual information
signal relative to the independent $128\,{\rm Mpc}$ simulation, even
around this characteristic scale. Instead, their MI amplitudes are
modestly suppressed, particularly near the maximum, although the
overall scale dependence and peak location remain comparable. This
indicates that the additional long-wavelength modes present in the
larger parent simulations do not produce a systematic enhancement of
the measured MI within the extracted subvolumes. The modest amplitude
differences nevertheless suggest that the parent simulation volume may
have a secondary influence, possibly through differences in the
large-scale fluctuation content. More importantly, the persistence of
the maximum at approximately the same voxel size across all three
measurements supports the interpretation that its location is governed
primarily by the effective analysis volume and the associated
finite-sampling limitations of the mutual information estimator,
rather than by the long-wavelength modes available in the parent
simulations.

This result implies that the location and amplitude of the
characteristic maximum are strongly influenced by the effective number
of voxels entering the estimator, $N_{\rm
  vox}=\left(\frac{L}{d}\right)^3$, rather than being determined
uniquely by the physical long-wavelength content of the parent
simulation. As $d$ increases, $N_{\rm vox}$ decreases rapidly, making
the estimation of the joint and marginal probability distributions
progressively more susceptible to finite-sample effects. The observed
maximum should therefore be regarded as a characteristic feature of
the present estimator in the large-$d$ regime, rather than as evidence
for a preferred physical coherence scale.

\begin{figure}[t]
    \centering
    \includegraphics[width=\columnwidth/2]{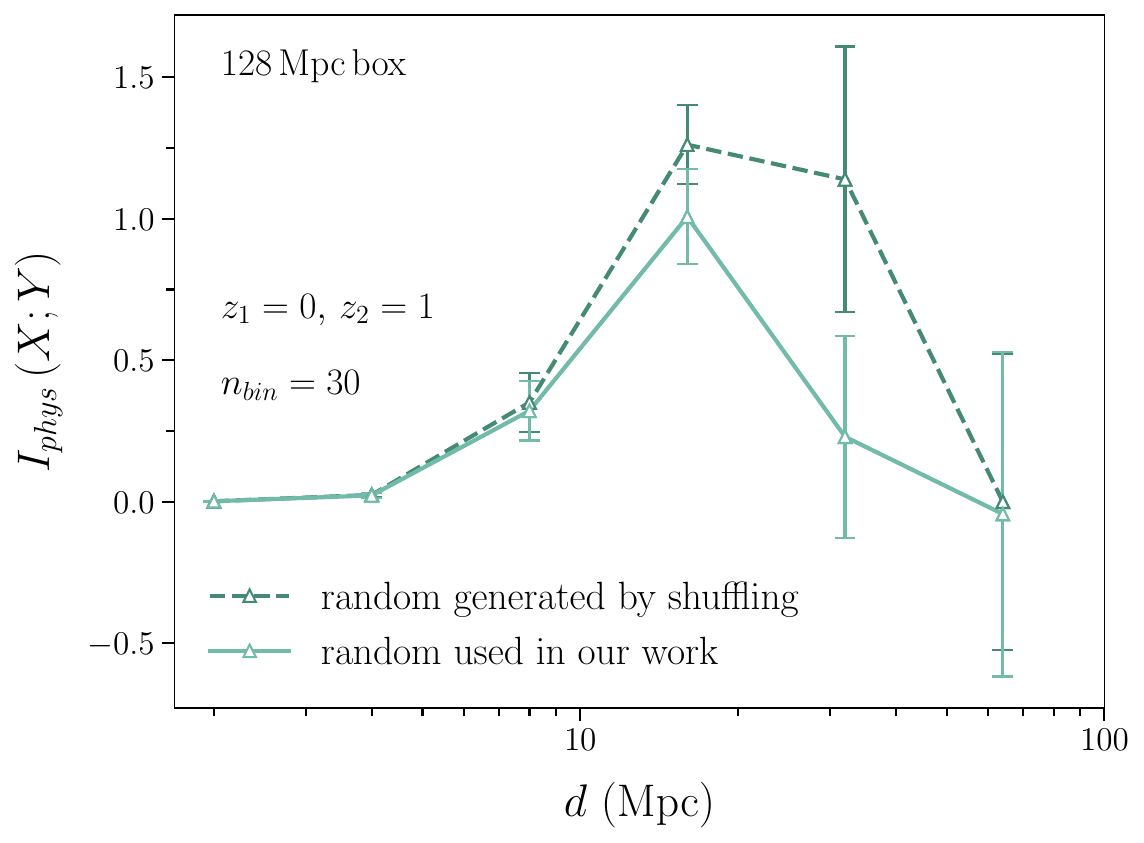}
    \caption{ \textbf{Effect of the reference baseline on the excess
        mutual information signal.}  Shared mutual information
      measured between the density fields at $z_1=0$ and $z_2=1$ in
      the $L=128~{\rm Mpc}$ simulation using two different baseline
      constructions. The null baseline (used throughout this work) is
      obtained from independent structureless random point
      distributions with the same particle number and box size as the
      simulation. The shuffled baseline is constructed by randomly
      permuting the voxel counts independently in each snapshot,
      preserving the one-point density distribution while removing all
      spatial correlations. In both cases, the physical mutual
      information is defined by subtracting the corresponding baseline
      mutual information from the measured mutual information of the
      simulation density fields. The two baseline choices yield
      qualitatively identical scale dependence, including the same
      characteristic peak location and large-scale decline.}
    \label{fig:new2}
\end{figure}

\subsection{Robustness to the choice of baseline}
\label{subsec:baseline}

Throughout this work, the physical mutual information associated with
gravitational evolution is quantified relative to a structureless
random baseline. Specifically, we define (\autoref{eq:smi})
\begin{equation}
I_{\rm phys}(d)=I_{\rm data}(d)-I_{\rm null}(d),
\label{eq:iphys_null}
\end{equation}
where $I_{\rm data}(d)$ is the mutual information measured from the
simulation snapshots and $I_{\rm null}(d)$ is the corresponding
mutual information measured between two independent random point
distributions with identical particle number and box size. This
baseline removes contributions arising from finite sampling,
voxelization, and estimator bias in the absence of any physical
correlations.

We also test whether our conclusions depend on the specific choice of
baseline. To address this point, we performed an additional analysis
using a shuffled-count baseline. For a given voxel size $d$,
the particle counts in all voxels are first measured from each
simulation snapshot. The voxel counts are then randomly shuffled
independently within each snapshot. This procedure preserves the
one-point distribution of voxel counts while destroying all spatial
correlations and all correspondence between the two epochs. The excess
mutual information is then defined as
\begin{equation}
I_{\rm phys}(d)=I_{\rm data}(d)-I_{\rm sh}(d),
\label{eq:iphys_shuffle}
\end{equation}
where $I_{\rm sh}(d)$ denotes the mutual information obtained from the
shuffled voxel-count distributions.

\autoref{fig:new2} compares the excess mutual information measured
using the two baselines for the $L=128~{\rm Mpc}$ simulation with
$n_{\rm bin}=30$. The overall behaviour remains qualitatively
unchanged. In both cases, the mutual information increases from small
voxel sizes, reaches a broad maximum at intermediate scales, and
declines at larger voxel sizes. Most importantly, the location of the
maximum remains unchanged.

The two baselines differ only quantitatively. The shuffled-count
baseline yields slightly larger residual mutual information,
particularly at large voxel sizes. This behaviour is expected because
the shuffled baseline preserves the non-Poisson nature of the voxel
count distribution and therefore provides a more conservative estimate
of the non-physical contribution to the measured mutual information.
Consequently, a somewhat larger excess signal remains after baseline
subtraction.

The close agreement between the two approaches demonstrates that the
measured scale dependence is not driven by the particular choice of
reference distribution. Although the random baseline of
\autoref{eq:iphys_null} is adopted throughout this work, the
shuffled-count test provides an important consistency check and
indicates that the inferred coherence signal reflects genuine
statistical relationships between the density fields rather than an
artifact of the baseline construction.

\subsection{Dependence on cosmic evolution}

\begin{figure*}[t]
    \centering
    \includegraphics[width=0.48\textwidth]{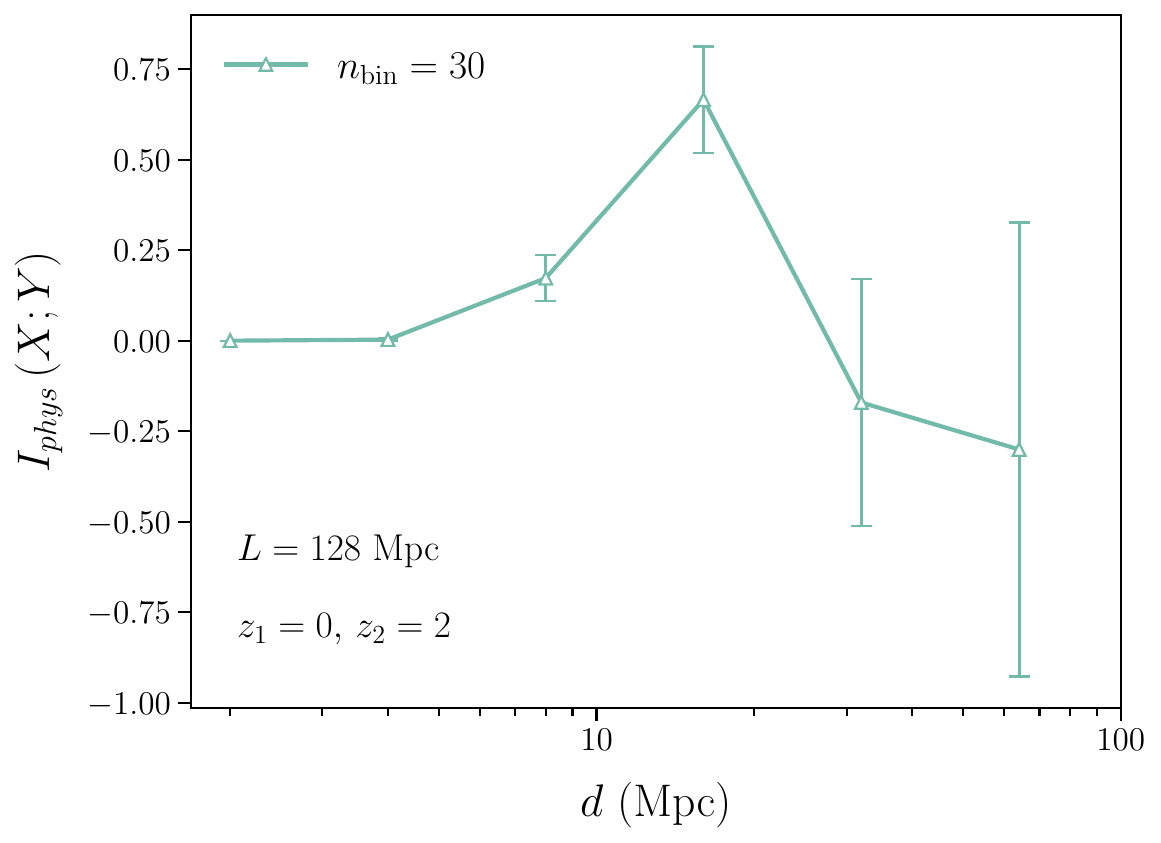}
    \hfill
    \includegraphics[width=0.48\textwidth]{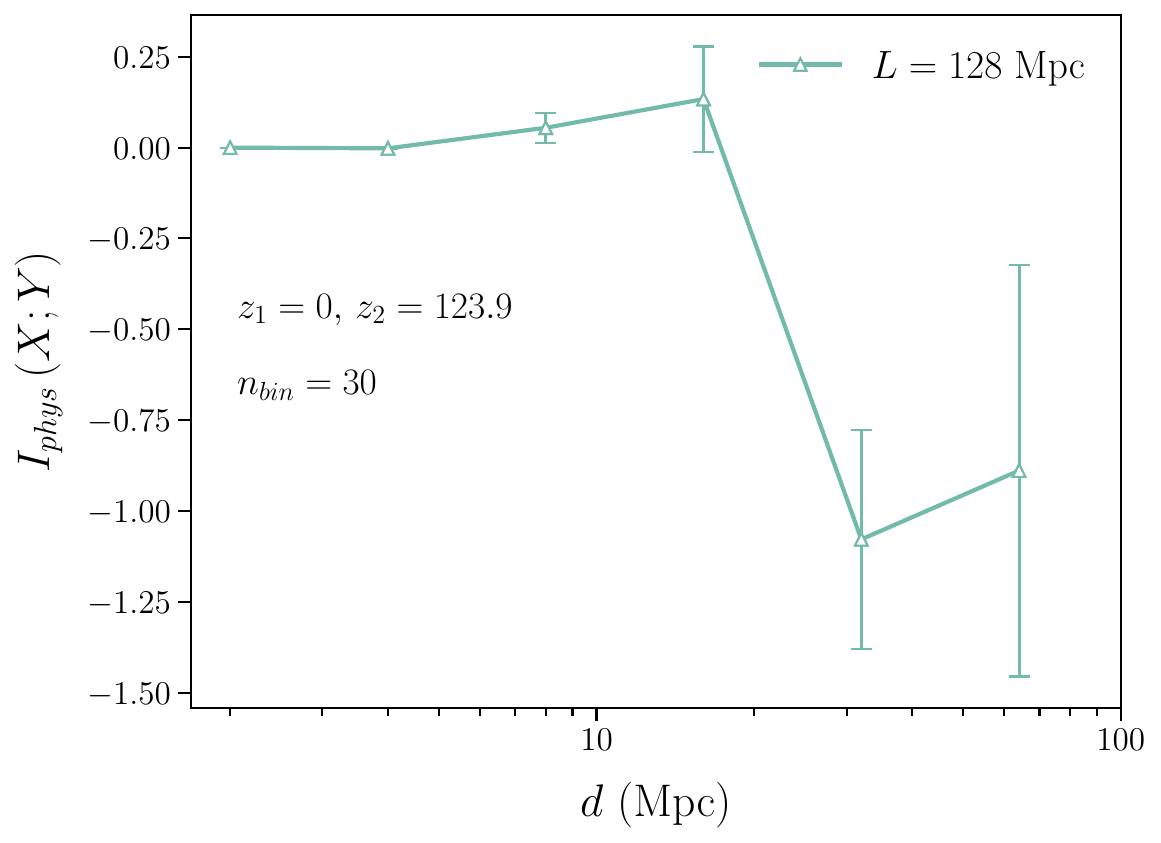}
    \caption{\textbf{Dependence of the shared mutual information on
        cosmic evolution.} Shared physical mutual information,
      $I_{\rm phys}(X;Y)$, as a function of voxel size $d$ for a
      $\Lambda$CDM $N$-body simulation with box size $L=128~{\rm Mpc}$
      and $n_{\rm bin}=30$.  \emph{Left:} comparison between density
      fields at $z_1=0$ and $z_2=2$.  \emph{Right:} comparison between
      the present-day density field ($z_1=0$) and the initial
      conditions ($z_2\simeq124$).  The shared mutual information is
      substantially reduced when the temporal separation between the
      snapshots increases. While the $z=0$ and $z=2$ density fields
      retain a measurable information overlap over a broad range of
      scales, the signal becomes much weaker when the present-day
      density field is compared directly to the initial
      conditions. This behaviour does not indicate a loss of
      information in the underlying gravitational dynamics, which
      remain deterministic, but rather a progressive reduction in the
      amount of information that can be recovered from the
      coarse-grained Eulerian density field at fixed spatial
      locations. Nonlinear mode coupling, bulk flows, mergers, and the
      redistribution of information across scales gradually transfer
      information into increasingly complex phase-space correlations,
      reducing the overlap between the initial and evolved density
      fields. The results demonstrate that mutual information provides
      a quantitative measure of how the observable imprint of
      gravitational memory becomes progressively diluted during
      nonlinear structure formation.}
    \label{fig:zigu4}
\end{figure*}

A key question is how the shared information measured by our
information-theoretic framework evolves as the temporal separation
between the density fields increases. To address this, we repeat the
analysis for a fixed simulation volume of $L=128~{\rm Mpc}$ using two
progressively larger redshift separations: $(z_1,z_2)=(0,2)$ and
$(0,124)$, where the latter corresponds to a comparison between the
present-day density field and the initial conditions of the
simulation. The resulting shared physical mutual information is shown
in \autoref{fig:zigu4}.

The comparison reveals a clear and physically intuitive trend.
For the $(z_1,z_2)=(0,2)$ pair, the shared mutual information remains
positive over a broad range of voxel sizes, indicating that a
substantial fraction of the large-scale organization of the density
field is still accessible through the coarse-grained Eulerian density
field at the two epochs. The overall scale dependence closely resembles
that seen in the previous analyses, with the signal rising from small
scales, attaining a maximum at intermediate scales, and declining at
larger voxel sizes.

In contrast, the shared mutual information measured between the
initial conditions and the present-day density field is dramatically
reduced. Although the signal remains positive, its amplitude is
substantially smaller across all voxel sizes, indicating that only a
limited fraction of the information present in the primordial density
field can be directly recovered from the evolved density field when
both are represented in a coarse-grained Eulerian framework. It may be
noted that at the largest voxel sizes, the baseline-subtracted mutual
information becomes slightly negative. This should not be interpreted
as negative mutual information. Rather, $I_{\rm phys}=I_{\rm
  data}-I_{\rm null}$ is a contrast relative to the null baseline, and
the negative values indicate that the estimated MI falls below this
baseline. At these large \(d\), the number of voxel samples, $N_{\rm
  vox}=(L/d)^3$, becomes small, while the physical MI for the
$z=0$--$z=124$ comparison is already extremely weak. The residual is
therefore increasingly dominated by finite-sample fluctuations and
estimator uncertainty. We regard the negative values as an indication
of the practical reliability limit of the present estimator at coarse
voxelization, rather than as a physical loss or reversal of
information.

This behaviour does not imply that gravitational evolution destroys
information. The dynamics of collisionless dark matter are
deterministic, and information is preserved in the full phase-space
description of the system. However, the information that remains
directly accessible through the Eulerian density field at fixed
spatial locations becomes progressively diluted over cosmic time.
Nonlinear mode coupling, bulk flows, mergers, shell crossing, and the
redistribution of information across scales all contribute to reducing
the overlap between the initial and evolved density fields. As matter
moves across voxel boundaries and structures undergo nonlinear
reorganization, information that was initially encoded in the density
field is transferred into increasingly complex phase-space
correlations that are not captured by a coarse-grained density
representation alone.

The comparison between the two redshift pairs therefore illustrates an
important aspect of the proposed framework. The shared mutual
information is sensitive not only to the spatial scale at which the
density field is examined, but also to the temporal depth over which
coherent information can be recovered from that field. Larger redshift
separations lead to systematically weaker information overlap,
providing a quantitative measure of how the observable imprint of
gravitational memory becomes progressively diluted during nonlinear
structure formation. This demonstrates that mutual information offers
a powerful and physically meaningful way to characterize the retention
of coherence across cosmic time, complementing conventional clustering
statistics that are insensitive to the shared information content
between different epochs.

\subsection{Dependence on cosmological parameters}

Having established the robustness of the mutual information framework
to voxelization choices, sample-size effects, and baseline
definitions, we now investigate its sensitivity to the underlying
cosmological model. While the previous subsections focused primarily
on the statistical reliability of the estimator, an equally important
question is whether the measured shared information carries
cosmological information and can therefore serve as a useful probe of
structure formation.

To address this question, we repeat the analysis using a set of
spatially flat $\Lambda$CDM simulations with identical box size
($L=128~{\rm Mpc}$) but different values of the cosmological density
parameters $(\Omega_m,\Omega_\Lambda)$. We consider four representative
models: $(0.1,0.9)$, $(0.315,0.685)$, $(0.5,0.5)$, and $(0.9,0.1)$.
For each model, we measure the shared physical mutual information
between the density fields at $z_1=0$ and $z_2=1$ using the same
analysis pipeline described above.

\begin{figure}[t]
    \centering
    \includegraphics[width=\columnwidth/2]{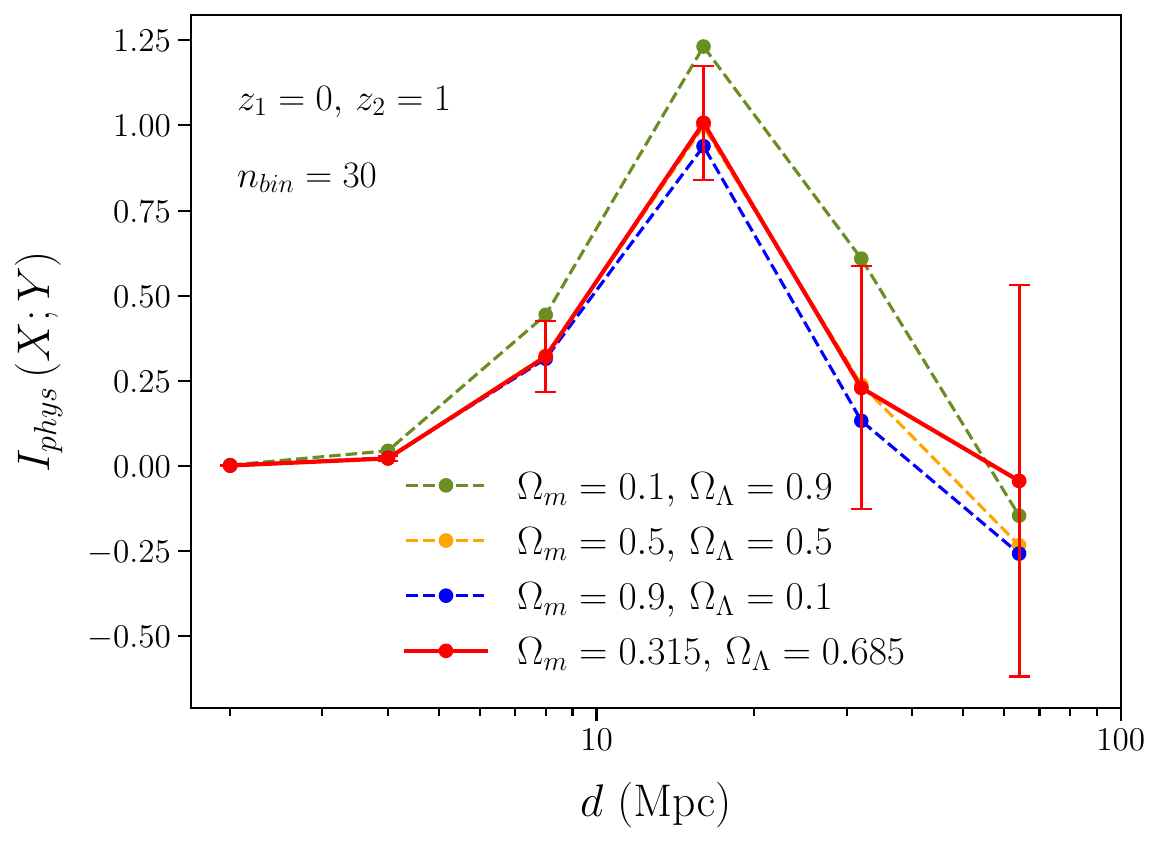}
    \caption{ \textbf{Dependence of shared physical mutual information
        on cosmological parameters.}  Shared physical mutual
      information, $I_{\rm phys}(X;Y)$, measured between the density
      fields at $z_1=0$ and $z_2=1$ for a set of flat $\Lambda$CDM
      cosmologies with different values of $(\Omega_m,\Omega_\Lambda)$
      in a simulation volume of size $L=128~{\rm Mpc}$.  While the
      overall scale dependence remains qualitatively similar across
      all models, the amplitude of the shared information varies
      systematically with cosmology.  Models with larger
      $\Omega_\Lambda$ retain higher mutual information between
      epochs, reflecting the slower nonlinear evolution and reduced
      mode coupling in such universes.  Conversely, matter-dominated
      cosmologies exhibit lower mutual information, indicating a more
      efficient redistribution of information across scales during
      structure formation.  The results demonstrate that the proposed
      information-theoretic framework is sensitive to the cosmological
      growth history and can discriminate between models through the
      amount of coherence retained in the evolved density field.  For
      clarity, $1\sigma$ error bars are shown for one representative
      cosmological model only.}
    \label{fig:zigu5}
\end{figure}

The results are shown in \autoref{fig:zigu5}. All models exhibit a
similar qualitative scale dependence. The shared mutual information is
small on very fine scales, increases as the density field is
progressively coarse-grained, reaches a maximum at intermediate
scales, and subsequently declines at large voxel sizes. As discussed
in the previous subsection, the location of this maximum is strongly
influenced by the effective sample size available to the estimator and
should therefore not be interpreted as a fundamental physical
coherence scale.

More importantly, \autoref{fig:zigu5} reveals a clear cosmological
dependence in the amplitude of the shared mutual information.  Models
with larger contributions from $\Lambda$ retain systematically higher mutual
information between epochs, whereas models dominated by matter show
lower amplitudes across a broad range of scales. This behaviour is
physically expected. A larger value of $\Omega_\Lambda$ suppresses the
late-time growth of structure, slowing nonlinear evolution and thereby
preserving a greater fraction of the statistical relationships between
density fields at different epochs. In contrast, higher matter
densities drive more efficient gravitational clustering, enhancing
nonlinear mode coupling and accelerating the redistribution of
information across scales. The resulting density fields therefore
retain less directly accessible coherence when viewed through a
coarse-grained Eulerian description.

These results demonstrate that the mutual information framework is not
merely sensitive to estimator properties but also responds to genuine
differences in cosmological evolution. The overall amplitude of the
shared information encodes how efficiently a cosmological model
preserves statistical coherence across cosmic time, providing a
complementary perspective on structure formation that is distinct from
traditional clustering statistics. While conventional measures such as
the power spectrum quantify the strength of density fluctuations at a
given epoch, the shared mutual information directly probes how much of
the information content of one cosmic state remains accessible in
another.

The dependence on cosmological parameters therefore highlights the
broader potential of this framework. Beyond serving as a tool for
quantifying coherence in the cosmic density field, it offers a new
information-theoretic diagnostic for comparing cosmological models and
studying how different growth histories preserve or erase statistical
memory during nonlinear gravitational evolution.

\subsection{Information retention across the cosmic web}
\label{subsec:environment}

A concrete application of the framework is to ask whether different
environments of the cosmic web retain different amounts of
statistically accessible information across cosmic time. In
particular, we ask whether the density fields associated with sheets,
filaments, and clusters exhibit different degrees of coherence between
$z=1$ and $z=0$, and how this information retention varies with
spatial scale. This provides a direct test of whether the statistical
memory of cosmic structure depends on environment.

We address this question using ten independent realizations of our
$\Lambda$CDM $N$-body simulations, each with box size $L=128~{\rm
  Mpc}$ and $262144$ particles. The cosmic web at $z=0$ and $z=1$ is
classified using the Hessian-based T-web method
\citep{hahn07,forero09}, with the underlying density field smoothed on
a scale of $2~{\rm Mpc}$, thereby separating the matter distribution
into sheets, filaments, and clusters. Because these environments occur
with different abundances, we construct matched samples by
downsampling the particles in each environment, separately at both
redshifts and across all realizations, to the minimum number available
for that environment. For each environment, the density fields at the
two redshifts are constructed using only the particles selected from
that environment at the respective epochs. The samples are matched in
particle number, and the mutual information is evaluated between the
resulting environment-specific density fields using the same
voxelization and estimation procedure described in
Section~2. Corresponding structureless random catalogues with
identical particle numbers are then generated, and the resulting
reference baseline is subtracted from the measured mutual information
to obtain the shared physical mutual information $I_{\rm phys}(X;Y)$
as a function of voxel size $d$ for each environment.

\begin{figure}[t]
    \centering
    \includegraphics[width=\columnwidth]{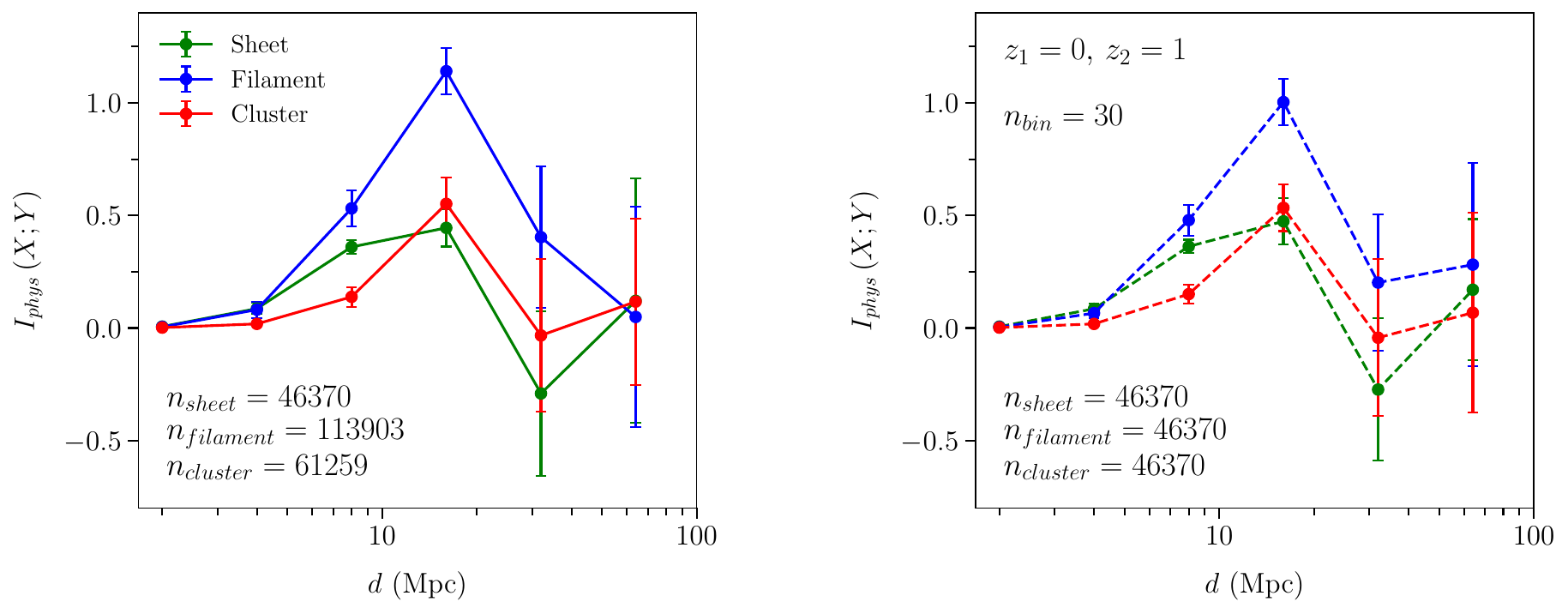}
    \caption{\textbf{Information retention across
          different cosmic-web environments.}  Shared physical mutual
        information, $I_{\rm phys}(X;Y)$, measured between the density
        fields at $z_1=0$ and $z_2=1$ in ten independent $\Lambda$CDM
        $N$-body realizations with $L=128~{\rm Mpc}$, after
        classifying the matter distribution into sheets, filaments,
        and clusters using the Hessian-based T-web method. The density
        field used for the classification is smoothed on a scale of
        $2~{\rm Mpc}$, and the mutual information is evaluated for
        $n_{\rm bin}=30$.  \emph{Left:} results using the matched
        samples available separately for each environment, with
        $N_{\rm sheet}=46370$, $N_{\rm filament}=113903$, and $N_{\rm
          cluster}=61259$.  \emph{Right:} results after imposing a
        common sample size, $N=46370$, on all three environments and
        regenerating the corresponding random catalogues with the same
        number of particles.  The environmental differences in the
        shared information persist after this strict sample-size
        matching, showing that the observed variation is not primarily
        caused by the unequal abundances of the three cosmic-web
        environments. The result demonstrates that mutual information
        can be combined with an independent cosmic-web classification
        to quantify how the statistically accessible density
        information retained across cosmic time depends on environment
        and spatial scale.}
    \label{fig:zigu6}
\end{figure}

The left panel of \autoref{fig:zigu6} reveals a clear environmental
dependence in the shared information. Filamentary environments retain
the largest amount of shared information between $z=1$ and $z=0$ over
a broad range of voxel sizes, while sheets and clusters exhibit lower
levels of information retention. The result indicates that the spatial
organization of matter at one epoch remains more strongly
statistically predictive of its later configuration in filaments than
in the other cosmic-web environments. Thus, the information-theoretic
measure distinguishes environmental differences in the evolution of
cosmic structure that are not expressed simply as a single global
measure of density-field coherence.

Because the three environments contain different numbers of particles,
however, their different sample sizes could in principle influence the
mutual information estimates. The numbers of particles in our matched
samples are $N_{\rm sheet}=46370$, $N_{\rm filament}=113903$, and
$N_{\rm cluster}=61259$. We therefore perform a stricter test in which
all three environments are restricted to the same number of particles,
$N=46370$, corresponding to the minimum count, and regenerate the
random catalogues using this common sample size. As shown in the right
panel of \autoref{fig:zigu6}, the environmental differences, including
the enhanced information retention in filaments, remain essentially
unchanged. The observed hierarchy is therefore not primarily a
consequence of unequal sample sizes.

\begin{figure*}
    \centering
    \includegraphics[width=0.85\textwidth]{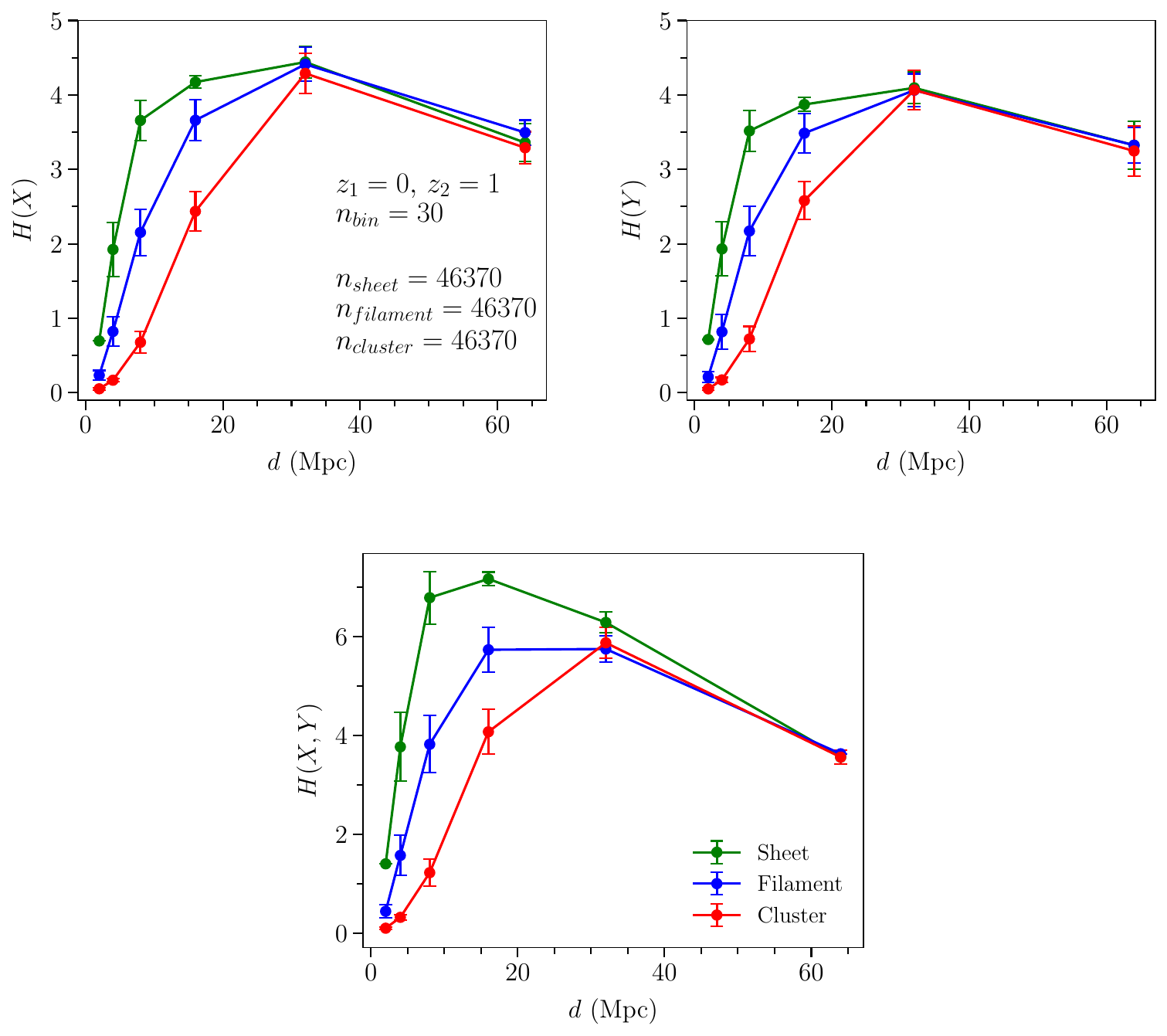}
    \caption{Marginal Shannon entropies $H(X)$ and $H(Y)$ (upper left
      and upper right panels, respectively) and joint entropy $H(X,Y)$
      (lower left panel) as functions of voxel size $d$ for sheets
      (green), filaments (blue), and clusters (red), measured between
      $z=1$ and $z=0$. All three environments are sampled with
      $N=46370$ particles to enable a direct comparison. Sheets
      generally exhibit the highest marginal and joint entropies at
      small and intermediate voxel sizes, followed by filaments and
      clusters, while the differences diminish at larger
      scales. Notably, filaments exhibit the largest shared mutual
      information (\autoref{fig:zigu6}) despite having lower marginal
      entropies than sheets, indicating that differences in
      information content alone cannot explain the observed
      environmental hierarchy. Error bars represent the $1\sigma$
      scatter across ten independent realizations.}
    \label{fig:zigu7}
\end{figure*}

The enhanced mutual information observed in filaments raises a further
question: could this result simply reflect differences in the
information content of the density fields associated with different
cosmic-web environments? To investigate this possibility, we compare
the marginal Shannon entropies, $H(X)$ and $H(Y)$, and the joint
entropy, $H(X,Y)$, for sheets, filaments, and clusters as functions of
voxel size, using the same particle number matched samples. As shown
in \autoref{fig:zigu7}, sheets generally exhibit the highest marginal
and joint entropies at small and intermediate voxel sizes, followed by
filaments and clusters, while these differences diminish at larger
scales. This hierarchy is consistent with the distinct spatial
distributions of matter across the cosmic web, with sheets occupying
extended two-dimensional structures, filaments forming elongated
networks, and clusters representing more strongly concentrated
configurations. Since mutual information is bounded by the marginal
entropies, $I(X;Y)\leq\min\{H(X),H(Y)\}$, a larger information content
can, in principle, accommodate a greater amount of shared
information. Interestingly, however, filaments exhibit the largest
mutual information despite not possessing the highest marginal
entropy. This demonstrates that the observed environmental hierarchy
cannot be attributed simply to a greater information capacity in
filamentary density fields.

The distinction between information content and shared information
points to a possible role of gravitational evolution in shaping the
statistical memory of different cosmic-web environments. While
marginal entropy characterizes the information content of a density
field at an individual epoch, mutual information also depends on the
statistical relationship between its configurations at different
epochs, as expressed by $I(X;Y)=H(X)+H(Y)-H(X,Y)$. The comparatively
lower mutual information in sheets, despite their higher entropies,
may reflect the continued anisotropic collapse and redistribution of
matter into filaments, which alter their spatial organization over
time. Clusters, on the other hand, undergo more advanced nonlinear
evolution involving mergers, shell crossing, and phase-space
mixing. Filaments occupy an intermediate evolutionary regime in which
their extended spatial organization may remain comparatively
persistent across cosmic epochs. Their enhanced mutual information may
therefore reflect a combination of sufficient information content and
stronger inter-epoch statistical dependence, rather than information
capacity alone. Although identifying the precise dynamical mechanisms
requires further investigation, these results suggest that the
statistical memory of the cosmic density field is shaped not only by
how information is distributed across environments, but also by how
gravitational evolution reorganizes that information over time.

\subsection{Practical guidelines for mutual information measurements}
\label{subsec:guidelines}

Our tests provide several practical guidelines for applying the
proposed information-theoretic framework to finite cosmological
datasets. The mutual information should always be evaluated after
explicitly accounting for the finite-sample contribution, since the
number of independent voxels decreases rapidly as the voxel size is
increased.  The density field should therefore be analysed over a
range of voxel sizes rather than at a single arbitrarily chosen scale,
and the convergence of the inferred signal should be checked against
the available number of voxels, $N_{\rm vox}=(L/d)^3$.

The choice of density binning is comparatively less restrictive. We
find that varying the number of occupied density bins over the range
tested in this work leaves the qualitative scale dependence of the
shared mutual information essentially unchanged. A fixed, moderate
number of bins, such as the $n_{\rm bin}=30$ adopted for most of our
analysis, therefore provides a practical choice, but convergence tests
with alternative bin numbers are recommended for new applications.

The reference baseline should be constructed using the same
voxelization, sampling, and number of objects as the data. In this
work, we use independent, structureless random point distributions as
the primary null baseline throughout. We find that an alternative
shuffled-count baseline, which preserves the one-point voxel-count
distribution while removing spatial correlations, gives the same
qualitative scale dependence and a slightly more conservative estimate
of the non-physical contribution, particularly at large voxel sizes.
We therefore recommend testing the baseline dependence when applying
the method to datasets with substantially different sampling
properties.

Most importantly, the voxelization scale should not be interpreted as
an independently determined physical coherence scale. Increasing $d$
simultaneously changes the spatial resolution and the statistical
sample size available to the estimator. Our subvolume tests show that
the characteristic maximum in the measured mutual information tracks
the effective size of the analysed volume, indicating that it is
primarily associated with finite-sample properties of the estimator.
For the present voxel-based implementation, scales approaching $d\sim
L/8$ should consequently be treated as an empirical reliability
threshold rather than as a physical boundary. In practical
applications, physical conclusions should be drawn from scales well
within this range and should be accompanied by convergence tests in
voxel size, binning, and baseline choice.

Overall, these considerations suggest a simple strategy: first
determine the scientifically relevant range of spatial scales, then
verify that a sufficient number of voxels samples the associated
probability distributions, adopt a reproducible baseline matched to
the data sampling, and finally demonstrate stability of the inferred
mutual information signal to reasonable changes in binning and
voxelization. Such tests are essential for distinguishing genuine
information retention from estimator-induced structure and make the
method reproducible across simulations and observational datasets.


\section{Conclusions}
\label{sec:conclusion}

We have introduced an information-theoretic framework for quantifying
the degree of coherence retained in the cosmic density field across
cosmic time. By measuring the mutual information between density
fields at different redshifts and subtracting an appropriate reference
baseline, the proposed approach isolates the excess statistical
dependence associated with gravitational evolution. Unlike
conventional two-point statistics, which quantify clustering
amplitudes at a single epoch, mutual information provides a direct
measure of the shared information between density fields and is
sensitive to nonlinear and higher-order correlations. This makes it a
natural tool for investigating how cosmic structure preserves,
redistributes, and transforms information during gravitational
evolution.

Applying this framework to a suite of cosmological $N$-body
simulations, we find that the shared mutual information exhibits a
strong and systematic dependence on spatial scale. The signal is
suppressed on very small scales, increases as the density field is
coarse-grained, and reaches a maximum before declining at larger
scales. Through a series of tests involving different binning
schemes, simulation volumes, alternative baseline constructions, and
subvolume analyses, we demonstrate that the location of this maximum
is governed primarily by the interplay between voxelization and the
finite sample size available to the mutual information estimator. The
observed peak should therefore not be interpreted as a fundamental
physical coherence scale of the Universe. Instead, it provides an
empirical indication of the scale beyond which the present estimator
becomes increasingly affected by finite-sampling limitations.

While the precise location of the maximum could be estimator
dependent, the underlying mutual information framework remains
physically meaningful. The analysis reveals clear differences in the
amount of shared information retained between different cosmic epochs
and across different cosmological models. In particular, the mutual
information between the initial conditions and the present-day density
field is found to be positive but extremely small, indicating that
although gravitational evolution is deterministic, the information
directly accessible through the coarse-grained Eulerian density field
at fixed spatial locations becomes progressively diluted over cosmic
time.  Nonlinear mode coupling, bulk flows, mergers, and the
redistribution of information across scales all contribute to reducing
the overlap between early- and late-time density fields.

The cosmological dependence of the signal further demonstrates that
mutual information can serve as a sensitive probe of structure-growth
history. Different cosmological models retain different amounts of
shared information between epochs, reflecting differences in the rate
at which nonlinear gravitational evolution redistributes information
through the density field. This suggests that information-theoretic
measures may provide a useful complement to traditional clustering
statistics when comparing cosmological models or characterizing the
growth of structure.

An important physical outcome of the present analysis is that mutual
information provides a direct measure of the predictability of the
density field across cosmic time at fixed spatial locations. The
measured $I_{\rm phys}(X;Y)$ therefore quantifies the amount of
coarse-grained density information that remains statistically encoded
in the evolved field, after accounting for contributions associated
with finite sampling, discretization, and the adopted reference
distribution. Its dependence on scale and amplitude provides a
complementary perspective to conventional clustering statistics:
rather than asking only how strongly matter clusters at a given epoch,
it asks how much of the spatial organization of the density field
remains statistically identifiable from another epoch. This makes the
framework a useful diagnostic of information retention and
redistribution during nonlinear gravitational evolution.

The present analysis is performed in an Eulerian framework, comparing
density fields at the same comoving spatial locations. A reduction in
Eulerian mutual information should therefore not be interpreted
automatically as irreversible loss of physical
information. Large-scale flows, displacements, mergers, and nonlinear
mode coupling can transport information between spatial locations and
redistribute it across scales and degrees of freedom, reducing the
information accessible through the Eulerian density field even when
the underlying matter distribution remains dynamically
connected. Extending the framework to a Lagrangian description, in
which density fields are compared at the same initial matter elements
across cosmic time, would provide a complementary measure of
information retention. A systematic comparison of Eulerian and
Lagrangian mutual information could therefore distinguish information
redistribution by bulk flows from a genuine loss of predictability,
offering a new quantitative probe of how information flows through
nonlinear structure formation.

A particularly promising application demonstrated in this work is the
ability to resolve information retention across different environments
of the cosmic web. We find that filamentary environments retain more
statistically accessible shared information between $z=1$ and $z=0$
than sheets and clusters, a hierarchy that persists after controlling
for differences in particle numbers. Interestingly, our additional
entropy analysis reveals that sheets generally possess higher marginal
and joint entropies than filaments, demonstrating that the enhanced
mutual information in filaments cannot be attributed simply to greater
marginal information content. Instead, the result suggests that the
statistical memory of the cosmic density field depends not only on the
information available within each environment but also on how its
spatial organization evolves under gravity. The enhanced coherence in
filaments may reflect their intermediate stage of gravitational
evolution, where extended structures remain comparatively persistent
without undergoing the advanced nonlinear collapse characteristic of
dense clusters. Although the precise dynamical origin of this
hierarchy requires further investigation, these findings demonstrate
that mutual information provides a sensitive quantitative probe of the
interplay between information content, gravitational evolution, and
environmental dependence in cosmic structure formation.

The mutual information developed here is not intended to replace the
power spectrum, correlation function, or other conventional measures
of spatial clustering. Rather, it addresses a different question by
quantifying how statistically predictable the density field at one
epoch remains from that at another. In this sense, the framework
provides a complementary information-theoretic perspective on
structure formation and opens several promising directions for future
research. Information-theoretic measures of coherence can be used to
investigate the emergence of statistical homogeneity, the
environmental dependence of galaxy formation and evolution, the
information content of large-scale surveys, and the fidelity of
cosmological simulations. The same framework can be extended beyond
the density field to velocity fields, tidal environments, halo
populations, and multi-tracer observables, and can naturally be
applied to quantify the memory of initial conditions and the
dependence of information retention on cosmological parameters. Such
extensions could provide a unified language for studying how
information is stored, redistributed, and retained in nonlinear
gravitational systems, establishing mutual information as a
complementary tool for connecting structure formation with the
information content of the observable Universe.

By introducing a quantitative measure of shared information in the
cosmic density field and carefully assessing its statistical
properties, this work establishes a foundation for a broader
information-theoretic description of large-scale structure. We hope
that these methods will stimulate further exploration of how
information is preserved, transformed, and ultimately encoded in the
evolving Universe.

\section{Acknowledgements}
The authors gratefully acknowledge the insightful comments and useful
suggestions from an anonymous reviewer which have significantly
improved the manuscript. BP acknowledges financial support from
Government of India through the project ANRF/ARG/2025/000535/PS. BP
also acknowledges IUCAA, Pune for providing support through
associateship programme. AN acknowledges the financial support from
the Department of Science and Technology (DST), Government of India
through an INSPIRE fellowship.

\bibliographystyle{JHEP}
\bibliography{mi.bib}

\label{lastpage}

\end{document}